\documentclass[twoside,twocolumn,9pt]{article}
\usepackage{extsizes}
\usepackage[super,sort&compress,comma]{natbib} 
\usepackage[version=3]{mhchem}
\usepackage[left=1.5cm, right=1.5cm, top=1.785cm, bottom=2.0cm]{geometry}
\usepackage{balance}
\usepackage{mathptmx}
\usepackage{sectsty}
\usepackage{graphicx} 
\usepackage{lastpage}
\usepackage[format=plain,justification=justified,singlelinecheck=false,font={stretch=1.125,small,sf},labelfont=bf,labelsep=space]{caption}
\PassOptionsToPackage{usenames,dvipsnames}{xcolor}
\usepackage{xcolor}
\usepackage{pdfpages}
\usepackage{float}
\usepackage{fancyhdr}
\usepackage{fnpos}
\usepackage[english]{babel}
\addto{\captionsenglish}{%
  
}
\usepackage{array}
\usepackage{droidsans}
\usepackage{charter}
\usepackage[T1]{fontenc}
\usepackage{setspace}
\usepackage[compact]{titlesec}
\usepackage{hyperref}

\usepackage{booktabs}
\usepackage{makecell}
\usepackage{pifont}
\newcommand{\cmark}{\ding{51}}
\newcommand{\xmark}{\ding{55}}
\renewcommand{\cmark}{\textcolor{green!60!black}{\ding{51}}}
\renewcommand{\xmark}{\textcolor{red!75!black}{\ding{55}}} 

\definecolor{cream}{RGB}{222,217,201}

\begin{document}

\pagestyle{fancy}
\thispagestyle{plain}
\fancypagestyle{plain}{
\renewcommand{\headrulewidth}{0pt}
}

\makeFNbottom
\makeatletter
\renewcommand\LARGE{\@setfontsize\LARGE{15pt}{17}}
\renewcommand\Large{\@setfontsize\Large{12pt}{14}}
\renewcommand\large{\@setfontsize\large{10pt}{12}}
\renewcommand\footnotesize{\@setfontsize\footnotesize{7pt}{10}}
\makeatother

\renewcommand{\thefootnote}{\fnsymbol{footnote}}
\renewcommand\footnoterule{\vspace*{1pt}%
\color{cream}\hrule width 3.5in height 0.4pt \color{black}\vspace*{5pt}} 
\setcounter{secnumdepth}{5}

\makeatletter 
\renewcommand\@biblabel[1]{#1}            
\renewcommand\@makefntext[1]%
{\noindent\makebox[0pt][r]{\@thefnmark\,}#1}
\makeatother 
\renewcommand{\figurename}{\small{Fig.}~}
\renewcommand{\tablename}{\small{Tab.}~}
\sectionfont{\sffamily\Large}
\subsectionfont{\normalsize}
\subsubsectionfont{\bf}
\setstretch{1.125} 
\setlength{\skip\footins}{0.8cm}
\setlength{\footnotesep}{0.25cm}
\setlength{\jot}{10pt}
\titlespacing*{\section}{0pt}{4pt}{4pt}
\titlespacing*{\subsection}{0pt}{15pt}{1pt}

\fancyfoot{}
\fancyfoot[LO,RE]{\vspace{-7.1pt}\includegraphics[height=9pt]{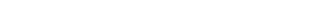}}
\fancyfoot[CO]{\vspace{-7.1pt}\hspace{13.2cm}\includegraphics{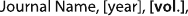}}
\fancyfoot[CE]{\vspace{-7.2pt}\hspace{-14.2cm}\includegraphics{head_foot/RF}}
\fancyfoot[RO]{\footnotesize{\sffamily{1--\pageref{LastPage} ~\textbar  \hspace{2pt}\thepage}}}
\fancyfoot[LE]{\footnotesize{\sffamily{\thepage~\textbar\hspace{3.45cm} 1--\pageref{LastPage}}}}
\fancyhead{}
\renewcommand{\headrulewidth}{0pt} 
\renewcommand{\footrulewidth}{0pt}
\setlength{\arrayrulewidth}{1pt}
\setlength{\columnsep}{6.5mm}
\setlength\bibsep{1pt}

\makeatletter 
\newlength{\figrulesep} 
\setlength{\figrulesep}{0.5\textfloatsep} 

\newcommand{\topfigrule}{\vspace*{-1pt}%
\noindent{\color{cream}\rule[-\figrulesep]{\columnwidth}{1.5pt}} }

\newcommand{\botfigrule}{\vspace*{-2pt}%
\noindent{\color{cream}\rule[\figrulesep]{\columnwidth}{1.5pt}} }

\newcommand{\dblfigrule}{\vspace*{-1pt}%
\noindent{\color{cream}\rule[-\figrulesep]{\textwidth}{1.5pt}} }

\makeatother

\twocolumn[
  \begin{@twocolumnfalse}
{\includegraphics[height=30pt]{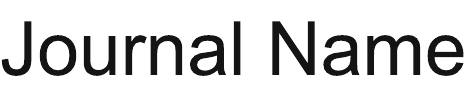}\hfill\raisebox{0pt}[0pt][0pt]{\includegraphics[height=55pt]{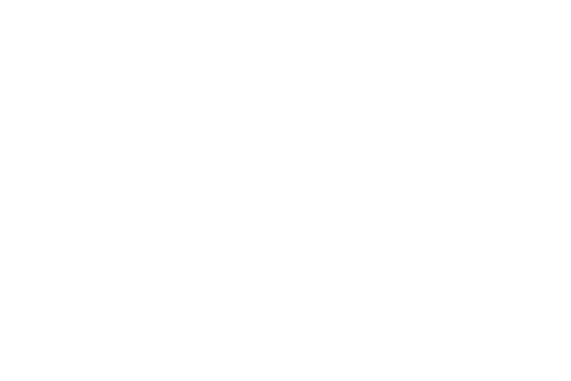}}\\[1ex]
\includegraphics[width=18.5cm]{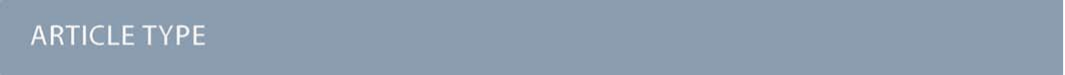}}\par
\vspace{1em}
\sffamily
\begin{tabular}{m{4.5cm} p{13.5cm} }

\includegraphics{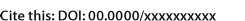} & \noindent\LARGE{\textbf{Van der Waals devices for surface-sensitive experiments$\dag$}} \\
\vspace{0.3cm} & \vspace{0.3cm} \\

 & \noindent\large{Nicolai~Taufertshöfer,\textit{$^{a,b}$} Corinna~Burri,\textit{$^{a,c}$} Rok~Venturini,\textit{$^{a}$} Iason~Giannopoulos,\textit{$^{a}$} \mbox{Sandy~Adhitia~Ekahana,\textit{$^{a}$}} Enrico~Della~Valle,\textit{$^{a,c}$} Anže~Mraz,\textit{$^{d,e}$} Yevhenii~Vaskivskyi,\textit{$^{d,e}$} Jan~Lipic,\textit{$^{d}$} Alexei~Barinov,\textit{$^{f}$} Dimitrios~Kazazis,\textit{$^{a}$} Yasin~Ekinci,\textit{$^{a}$} Dragan~Mihailovic,$^{\ast}$\textit{$^{d,e,g}$} \mbox{and~Simon~Gerber$^{\ast}$\textit{$^{a}$}}} \\

\includegraphics{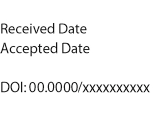} & \noindent\normalsize{
In-operando characterization of van der Waals~(vdW) devices using surface-sensitive methods provides critical insights into phase transitions and correlated electronic states. Yet, integrating vdW~materials in functional devices while maintaining pristine surfaces is a key challenge for combined transport and surface-sensitive experiments. Conventional lithographic techniques introduce surface contamination, limiting the applicability of state-of-the-art spectroscopic probes. We present a stencil lithography-based approach for fabricating vdW~devices, producing micron-scale \mbox{electrical} contacts, and exfoliation in ultra-high vacuum. The resist-free patterning method utilizes a shadow mask to define electrical contacts and yields thin flakes down to the single-layer regime via gold-assisted exfoliation. As a demonstration, we fabricate devices from \mbox{1\textit{T}--TaS\textsubscript{2}} flakes, achieving reliable contacts for application of electrical pulses and resistance measurements, \mbox{as well} as clean surfaces allowing for angle-resolved photoemission spectroscopy. The approach provides a platform for studying the electronic properties of vdW~systems with surface-sensitive probes in well-defined device geometries.
} \\

\end{tabular}

 \end{@twocolumnfalse} \vspace{0.6cm}

  ]

\renewcommand*\rmdefault{bch}\normalfont\upshape
\rmfamily
\section*{}
\vspace{-1cm}


\footnotetext{\textit{$^{a}$~PSI Center for Photon Science, Paul Scherrer Institute, 5232 Villigen PSI, Switzerland. E-mail: simon.gerber@psi.ch}}
\footnotetext{\textit{$^{b}$~Institute for Quantum Electronics, ETH Zurich, 8093 Zurich, Switzerland.}}
\footnotetext{\textit{$^{c}$~Laboratory for Solid State Physics and Quantum Center, ETH Zurich, 8093 Zurich, Switzerland.}}
\footnotetext{\textit{$^{d}$~Department of Complex Matter, Jožef Stefan Institute, 1000 Ljubljana, Slovenia. \mbox{E-mail}: dragan.mihailovic@ijs.si}}
\footnotetext{\textit{$^{e}$~CENN Nanocenter, 1000 Ljubljana, Slovenia.}}
\footnotetext{\textit{$^{f}$~Elettra - Sincrotrone Trieste, 34149 S.C.p.A., Basovizza (TS), Italy.}}
\footnotetext{\textit{$^{g}$~Faculty of Mathematics and Physics, University of Ljubljana, 1000 Ljubljana, \mbox{Slovenia}.}}

\footnotetext{\dag~Supplementary Information available: See DOI: \href{https://doi.org/10.5281/zenodo.15482369}{10.5281/zenodo.15482369}.}



\section{Introduction}
Van der Waals~(vdW) materials provide a rich playground for \mbox{exploring} correlated electronic states and phase transitions in two-dimensional systems, with applications in next-generation electronic devices based on thin vdW flakes. They enable the exploration of novel quantum states with unprecedented control via gating and pulse application \cite{caoUnconventionalSuperconductivityMagicangle2018, radisavljevicSinglelayerMoS2Transistors2011, yeSuperconductingDomeGateTuned2012, stojchevskaUltrafastSwitchingStable2014, venturiniElectricallyDrivenNonvolatile2024}. While transport measurements provide insight into the electronic properties of such \mbox{devices}, spectroscopic tools are necessary to directly probe the electronic band structure. \mbox{Combining} these approaches remains challenging as conventional lithographic device fabrication methods introduce surface contamination or modify the surface, preventing the use of state-of-the-art surface-sensitive tools, such as scanning tunneling microscopy~\cite{yinProbingTopologicalQuantum2021} or angle-resolved photoemission spectroscopy~(ARPES)~\cite{sobotaAngleresolvedPhotoemissionStudies2021}. Thus, a critical challenge in fabricating devices for surface-sensitive experiments involving \mbox{vdW~materials} is achieving three aspects simultaneously: (i)~obtaining thin, uniform flakes down to monolayers, (ii)~integrating electrical contacts at the micron-scale for well-defined device geometries, and (iii)~maintaining pristine surface quality through \textit{in-situ} \mbox{exfoliation} in \mbox{ultra-high} vacuum~(UHV).

\begin{figure*}
  \centering
  \includegraphics[width=\linewidth, trim=0cm 18.29cm 0cm 0cm, clip]{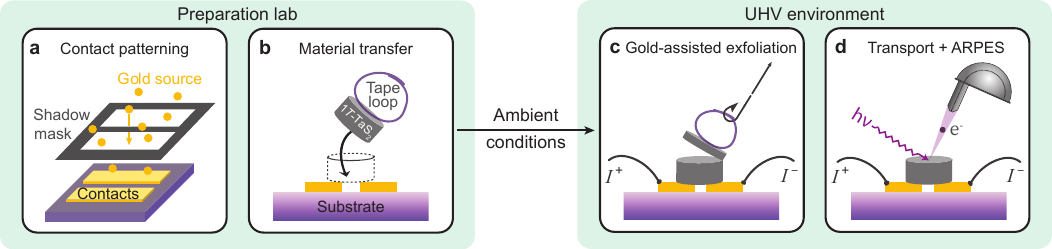}
  \caption{Schematic of the vdW device fabrication process using a stencil for contact patterning and gold-assisted exfoliation. \textbf{a} Au evaporation through a shadow mask, defining the contact geometry. \textbf{b} Bulk vdW material on a tape loop is transferred to the pre-patterned contacts immediately after Au~evaporation. \textbf{c} Gold-assisted exfoliation in UHV. \textbf{d} Integration of surface-sensitive techniques, \textit{e.g.} ARPES, and \textit{in-situ} transport measurements.}
  \label{fig:1}
\end{figure*}

A variety of fabrication methods have been developed to integrate vdW materials into devices, but most approaches satisfy only two out of the three requirements above (see Tab.~\ref{tab:methods_comparison}). Fabrication methods involving standard dry pick-up transfer techniques with viscoelastic polymer stamps, \mbox{\textit{e.g.} polydimethylsiloxane} (PDMS) \cite{castellanos-gomezDeterministicTransferTwodimensional2014}, are widely used for assembling vdW heterostructures, but they often leave polymer residues and water on the surface, which degrade the interface quality and limit compatibility with surface-sensitive techniques~\cite{ishigamiAtomicStructureGraphene2007, wangOneDimensionalElectricalContact2013}. While such \mbox{contamination} can in principle be removed, \textit{e.g.} from graphene devices by annealing at \mbox{$\approx 350\,^\circ$C \cite{lisiObservationFlatBands2021}}, this approach is viable only for certain ``robust'' vdW materials. In many cases, such treatment leads to irreversible structural changes. For instance, for the \mbox{prototypical} transition metal \mbox{dichalcogenide} material \mbox{1\textit{T}--TaS\textsubscript{2}}, \mbox{annealing} at \mbox{$\approx 325\,^\circ$C} induces a polytype transition to the semimetallic 2\textit{H} phase~\cite{givensThermalExpansionNbSe21977}. Alternative strategies to avoid polymer contamination include protective capping with hBN or graphene~\cite{masubuchiDryPickandflipAssembly2022} during transfer or the use of metalized SiN$_x$ membranes~\cite{wangCleanAssemblyVan2023}. Though, both approaches still require mild annealing and are difficult to integrate with patterned device architectures.

Cleaving bulk vdW crystals in UHV with pre-attached \mbox{contacts} has recently also been used in transport-ARPES studies on Ca\textsubscript{2}RuO\textsubscript{4} \cite{suenElectronicResponseMott2024} and \mbox{1\textit{T}--TaS\textsubscript{2}}~\cite{nitzavEmergenceFermisurfaceCurrentdriven2024}. This method exposes a pristine surface \textit{in-situ} but is limited to bulk materials, making it incompatible with micron-scale device designs.

\begin{table}[b!]
\renewcommand{\arraystretch}{1.3}
\setlength{\tabcolsep}{3pt}
\caption{Comparison of fabrication methods for vdW devices in terms of: (i)~monolayer/thin flake preparation, (ii)~micron-scale contact integration, and (iii)~pristine surface preservation. $^1$Clean surfaces can be achieved, but only by annealing. $^2$Patterning is possible, but compromises surface cleanliness.}
\label{tab:methods_comparison}
\centering
\begin{tabular}{p{6.5cm}@{\hspace{1.5em}}ccc}
\toprule
\textbf{Method} & \textbf{(i)} & \textbf{(ii)} & \textbf{(iii)} \\
\midrule
Dry pick-up transfer \cite{castellanos-gomezDeterministicTransferTwodimensional2014, lisiObservationFlatBands2021, masubuchiDryPickandflipAssembly2022, wangCleanAssemblyVan2023} & \cmark & \cmark & \xmark\rlap{$^1$} \\
UHV cleaving of bulk crystals \newline with pre-attached contacts \cite{suenElectronicResponseMott2024, nitzavEmergenceFermisurfaceCurrentdriven2024} & \xmark & \xmark & \cmark \\
Glovebox device assembly \cite{caoQualityHeterostructuresTwoDimensional2015, masubuchiAutonomousRoboticSearching2018, gantSystemDeterministicTransfer2020, patilPickupAssemblingChemically2024} & \cmark & \cmark & \xmark\rlap{$^1$} \\
UHV fabrication system \cite{guoUltrahighVacuumSystem2023} & \cmark & \xmark & \cmark \\
Gold-assisted exfoliation \cite{magdaExfoliationLargeareaTransition2015, desaiGoldMediatedExfoliationUltralarge2016, velickyMechanismGoldAssistedExfoliation2018, huangUniversalMechanicalExfoliation2020} & \cmark & \xmark\rlap{$^2$} & \cmark \\
\textbf{µ-stencil + gold-assisted exfoliation} \newline \textbf{(this work)} & \cmark & \cmark & \cmark \\
\bottomrule
\end{tabular}
\end{table}

Fabrication in inert atmospheres such as in N$_2$ or Ar~gloveboxes~\cite{caoQualityHeterostructuresTwoDimensional2015, masubuchiAutonomousRoboticSearching2018, gantSystemDeterministicTransfer2020, patilPickupAssemblingChemically2024}, or even under UHV~\cite{guoUltrahighVacuumSystem2023}, has been explored to exfoliate thin flakes and assemble heterostructures while minimizing oxidation. However, even within a controlled N$_2$~environment, water adsorption on the surface remains an issue, often necessitating an annealing step to obtain clean surfaces. Moreover, such fabrication setups require specialized equipment that is rarely available near UHV measurement facilities.

Finally, gold-assisted exfoliation~\cite{magdaExfoliationLargeareaTransition2015, desaiGoldMediatedExfoliationUltralarge2016, velickyMechanismGoldAssistedExfoliation2018, huangUniversalMechanicalExfoliation2020} allows for preparation of large-area monolayer flakes with clean surfaces by leveraging the strong interaction between freshly evaporated Au and chalcogen atoms. While Au is most commonly used due to its inert nature, it can be replaced by other metals such as Pd, Ni, Cu, and Ag, which exhibit similarly high adhesion through strong bonding to chalcogen atoms \cite{johnstonCanMetalsOther2022}. Importantly, the timing of exfoliation is critical: contaminants adsorb on Au within minutes, making the surface hydrophobic, which significantly reduces the monolayer yield~\cite{smithHydrophilicNatureClean1980, velickyMechanismGoldAssistedExfoliation2018}. This degradation cannot be reversed by cleaning or annealing, precluding post-processing steps such as metal lift-off. While the method yields high-quality flakes, device integration typically involves patterning steps after exfoliation, for example, by etching in potassium iodide solution~\cite{huangUniversalMechanicalExfoliation2020}.

To fulfill all three criteria~(i)–(iii) simultaneously (see Tab.~\ref{tab:methods_comparison}), we introduce a fabrication method that utilizes gold-assisted exfoliation with pre-patterned metallic contacts defined by evaporation through a stencil. This approach enables the fabrication of vdW devices with micron-scale contact geometries while preserving pristine surfaces through exfoliation in UHV. Unlike approaches that require integrated lithography clusters, our method allows for device fabrication in a standard laboratory environment and transport \mbox{to the} \mbox{experimental} facility, \textit{e.g.} synchrotrons or other dedicated UHV measurement setups, without specialized vacuum equipment. The fabrication process, illustrated in Fig.~\ref{fig:1}, consists of three main steps: First, Au contacts are deposited on a Si/SiO\textsubscript{2} substrate through a shadow mask, defining the device geometry. \mbox{Second,} immediately after venting the evaporation chamber, a freshly cleaved bulk crystal is transferred to the contacts using a tape loop. The device is then assembled on a sample holder and can be transported under ambient conditions. Third, exfoliation is performed inside the UHV measurement chamber by removing the tape, which cleaves the bulk crystal and exposes clean \mbox{vdW flakes} over the Au contacts.

As a demonstration of this technique, we fabricate two-terminal devices of \mbox{1\textit{T}--TaS\textsubscript{2}}, a prototypical vdW material due to its rich electronic phase diagram, \mbox{including} a Mott insulating low-temperature state~\mbox{\cite{fazekasElectricalStructuralMagnetic1979, perfettiTimeEvolutionElectronic2006, liggesUltrafastDoublonDynamics2018, butlerMottnessUnitcellDoubling2020, huaEffectInterlayerStacking2025}}, \mbox{different} charge-density wave (CDW) phases~\cite{wilsonChargedensityWavesSuperlattices1975, hughesChargeDensityWaves1976}, superconductivity upon pressure~\cite{siposMottStateSuperconductivity2008}, and a putative quantum spin liquid state~\cite{klanjsekHightemperatureQuantumSpin2017, chenSpectroscopicEvidencePossible2025}. Furthermore, \mbox{1\textit{T}--TaS\textsubscript{2}} exhibits a non-equilibrium metallic ``hidden'' state that can be reversibly induced by optical or electrical pulses~\mbox{\cite{stojchevskaUltrafastSwitchingStable2014,vaskivskyiFastElectronicResistance2016, stahlCollapseLayerDimerization2020, maklarCoherentLightControl2023, burriNondestructiveImagingBulk2024}}. This non-volatile state holds promise for use in the next generation of cryo-memory devices due to energy efficiency and switching speed~\mbox{\cite{vaskivskyiControllingMetaltoinsulatorRelaxation2015, venturiniUltraefficientResistanceSwitching2022, mrazChargeConfigurationMemory2022}}. 

The devices presented here are tailored for micro-beam ARPES \cite{cattelanPerspectiveApplicationSpatially2018, limRecentTechnicalAdvancements2024} in combination with transport measurements and application of electrical (switching) pulses. More broadly, the fabrication method enables studies of the electronic properties of thin vdW flakes and provides a platform for integrating surface-sensitive techniques with transport measurements across a wide range of material systems.

\begin{figure*}[t]
  \centering
  \includegraphics[width=\linewidth, trim=0cm 11cm 0cm 0cm, clip]{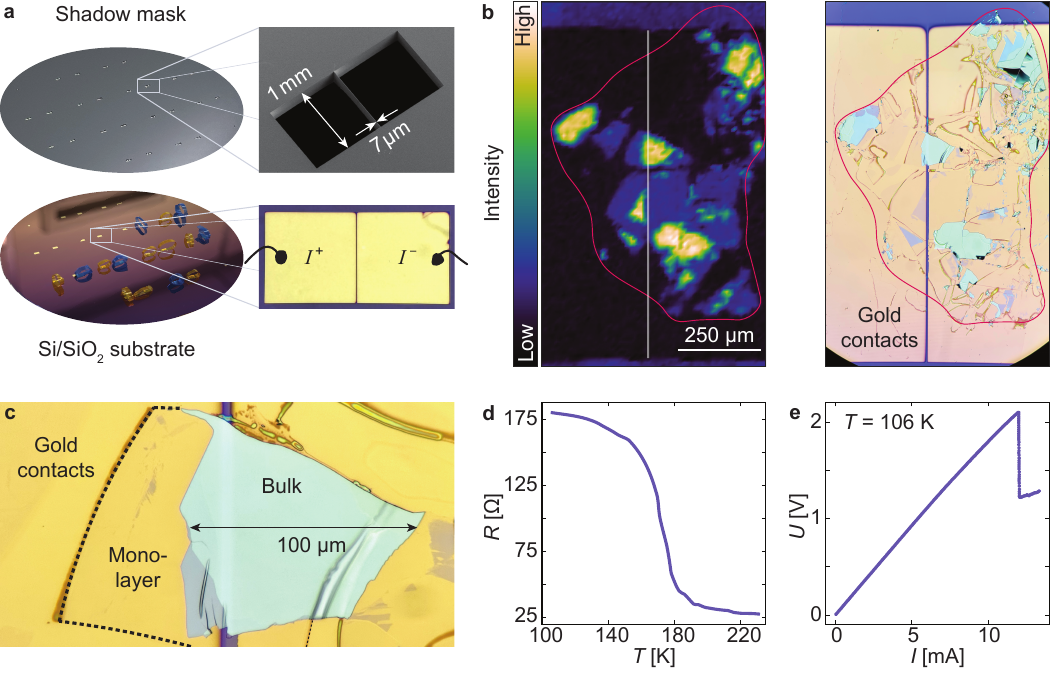}
  \caption{Fabrication and characterization of a two-terminal \mbox{1\textit{T}--TaS\textsubscript{2}} device. \textbf{a} Shadow mask with 24 copies of the device geometry (top) used to pattern a 4-inch \mbox{Si/SiO\textsubscript{2}} wafer with Au evaporation (bottom), on which tape loops with bulk \mbox{1\textit{T}--TaS\textsubscript{2}} crystals are then placed. \textbf{b} Ta~$4f$ core-level intensity map from spatially-resolved ARPES alongside an optical microscope image of the cleaved \mbox{1\textit{T}--TaS\textsubscript{2}} flakes. The outline of the bulk crystal is shown in red. \textbf{c} Optical microscope image of a \mbox{$45$-nm}-thick bulk flake bridging the gap between the contact pads, as well as an extended monolayer flake. \textbf{d}~\textit{In-situ} transport measurements of the \mbox{1\textit{T}--TaS\textsubscript{2}} device, showing the metal-insulator (\mbox{NC-CCDW}) transition upon cooling. \textbf{e} Pulsed \textit{IV}~measurements in the insulating CCDW state show electrical switching to the ``hidden'' state, marked by a voltage drop at $12\;$mA.}
  \label{fig:2}
\end{figure*}

\section{Results and discussion}

\subsection{Device fabrication}

Here, we outline the steps for the preparation of the vdW~\mbox{device}. Details of the shadow-mask fabrication are provided in the \mbox{Methods} section.
To benefit from the high adhesion of gold-assisted exfoliation, it is crucial to fabricate electrodes in a way that avoids contamination. This can be achieved through evaporation of Au through a shadow mask \cite{grayProductionFinePatterns1959, dunklebergerStencilTechniquePreparation1978, zhouSimpleFabricationMolecular2003, vazquez-menaResistlessNanofabricationStencil2015}. Such apertures can be made using various techniques, such as mechanical cutting, focused ion beam milling \cite{ozyilmazFocusedionbeamMillingBased2007, vazquez-menaMetallicNanowiresFull2008, batesEffectUsingStencil2013}, or etching into Si wafers~\mbox{\cite{burgerHighresolutionShadowmaskPatterning1996, parkPatterningParallelNanobridge2007, villanuevaEtchingSubmicrometerStructures2008}}, with the choice depending on the required feature size. For our approach, we employ laser lithography combined with deep reactive ion etching~(DRIE)~\cite{xuDeepreactiveIonEtching2024} to produce high-aspect ratio masks with well-defined micron-scale features (see Fig.~\ref{fig:2}a and the \mbox{Methods}). This stencil wafer is placed on the Si/SiO\textsubscript{2}~substrate during the deposition of Ti/Au contacts via electron beam \mbox{evaporation}. \mbox{After} the contact deposition, bulk \mbox{1\textit{T}--TaS\textsubscript{2}} crystals with typical dimensions of $1-3\;$mm in diameter are cleaved on tape loops, positioned over the gap between the Au pads, brought into contact with the substrate, and gently pressed to ensure adhesion. 

Due to rapid degradation of the Au surface in air, it is crucial that this is performed within $\approx5\;$min after venting the evaporation chamber to obtain a high yield of thin flakes \cite{velickyMechanismGoldAssistedExfoliation2018}. Next, \mbox{electrical} connections are made by attaching wires to the contact pads using Ag epoxy. A~fully assembled device is shown in Fig.~S1$\dag$. 

\begin{figure*}[t]
  \centering
  \includegraphics[width=\linewidth, trim=0cm 11.65cm 0cm 0cm, clip]{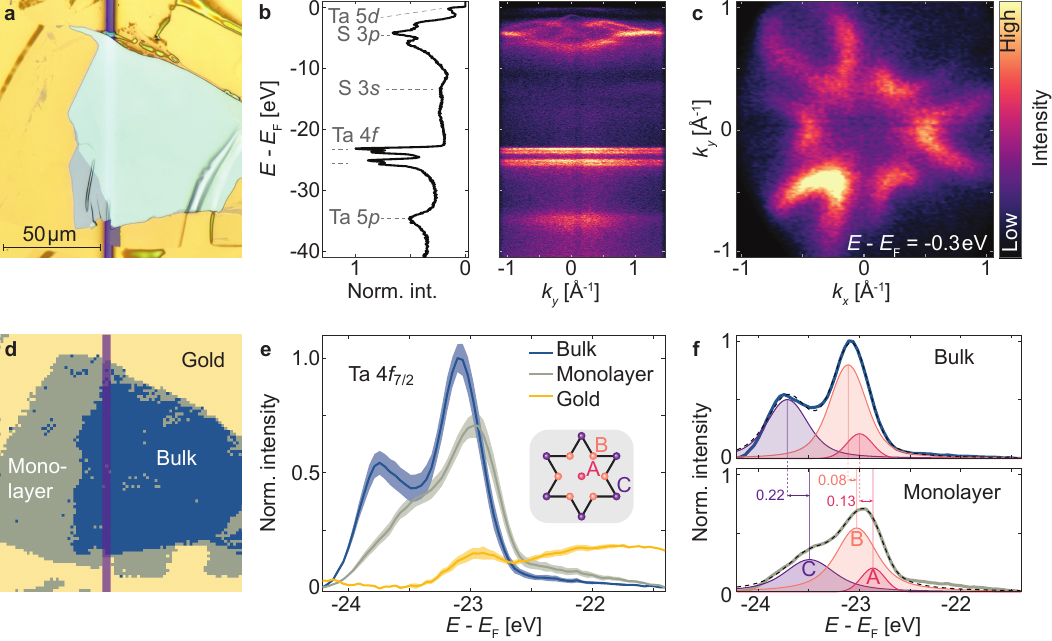}
  \caption{Spatially-resolved ARPES of a \mbox{1\textit{T}--TaS\textsubscript{2}} device. \textbf{a} Optical microscope image of the measured flake. \textbf{b}~ARPES spectrum with the angle-integrated elemental signatures (left) and the corresponding band structure (right). \textbf{c} Isoenergy contour integrated over a 0.1$\;$eV window centered at \mbox{$E-E_{\mathrm{F}} = -0.3\;$eV}. Streaks near normal emission in $k_y$ in panels b and c are artifacts from the analyzer’s deflector system. \textbf{d}~$k$-means clustering map (the spatial extent corresponds to panel a), identifying three distinct regions based on the Ta~$4f$ core-level spectrum. The contact gap is marked in purple. \textbf{e} Averaged Ta$\,4f_{7/2}$ core-level spectrum for all data points within the the clusters. The shaded area corresponds to the respective standard deviation. The inset depicts the star-of-David structure of Ta atoms which forms the building block of the CDW states of \mbox{1\textit{T}--TaS\textsubscript{2}} and features three distinct Ta species: the central site A, as well as the inner and outer ring site B and C, respectively. \textbf{f}~Fits of the Ta$\,4f_{7/2}$ bulk and monolayer spectra using three Voigt profiles, corresponding to the three inequivalent Ta environments. The sum of the three fitted components is depicted by dashed lines, whilst the experimental data is shown in bold.}
  \label{fig:3}
\end{figure*}

To obtain a contamination-free surface, the exfoliation process is performed later under UHV conditions by removing the tape loop with a mechanical manipulator. This process cleaves the vdW crystal, leaving flakes with freshly exposed surfaces on the contacts at random positions; only flakes that bridge the gap between the contact pads form functional devices. Electrical connectivity is tested and the flakes are visually inspected using an optical microscope, as well as in our ARPES demonstration the spectral intensity of the Ta~$4f$core levels, mainly to confirm flake dimensions, thickness, and surface homogeneity (see Fig.~\ref{fig:2}b). The vdW flakes, like the one shown in Fig.~\ref{fig:2}c, have lateral dimensions from tens to hundreds of microns. Atomic force microscopy (AFM) reveals thicknesses ranging from a monolayer up to $\approx 100\;$nm (see Fig.~S2$\dag$), enabling access to both few-layer and bulk properties. Such variation in thickness can \mbox{occur} even within a single macroscopic flake as local tearing and a non-uniform stress distribution during peeling, combined with the material's in-plane stiffness, can result in different exfoliation planes. In our experiments, we routinely observe mono-, bi-, and few-layer flakes. While the exfoliation process is inherently non-deterministic, the yield and thickness distribution of flakes can be influenced by tuning the Au surface properties—such as roughness and cleanliness—through parameters like evaporation rate, film thickness, adhesion layer, and time elapsed after breaking vacuum. This enables the selection of flakes with the desired thickness from the exfoliated set.

\subsection{Device characterization}
\label{sec:results}

\paragraph*{Transport measurements:}  
Figure \ref{fig:2}d shows the temperature dependence of the two-terminal \mbox{resistance} upon cooling after exfoliation at 250$\;$K, including the metal-insulator transition from the the nearly-commensurate (NC) to commensurate~(C) CDW~phase of \mbox{1\textit{T}--TaS\textsubscript{2}}. Large contact resistances on the order of several hundreds of $\Omega$ are typically observed when vdW \mbox{materials} are probed with metal contacts on their flat face~\cite{wangOneDimensionalElectricalContact2013}. However, we observe exceptionally low contact resistances on the order of $10\;\Omega$ in all our devices, \textit{e.g.} the two-terminal resistance of the device shown in Fig.~\ref{fig:2}c is $25\;\Omega$ at room temperature, which we attribute to the large contact area with covalent-like bonding at the \mbox{1\textit{T}--TaS\textsubscript{2}}$\,$/$\,$Au interface~\cite{huangUniversalMechanicalExfoliation2020}. The high-quality contacts allow for low-voltage state switching of the \mbox{1\textit{T}--TaS\textsubscript{2}} device, where electrical excitation induces a phase transition from the insulating, equilibrium CCDW phase to a metallic, non-thermal and metastable ``hidden'' state~\cite{vaskivskyiFastElectronicResistance2016}. This phase transition is characterized by a drop in resistance and associated with the disruption of long-range CCDW order. Recent studies suggest that the emergence of the hidden state is driven by out-of-plane restacking and the disappearance of interlayer dimerization of the CCDW~phase~\mbox{\cite{stahlCollapseLayerDimerization2020, maklarCoherentLightControl2023, burriNondestructiveImagingBulk2024, huaEffectInterlayerStacking2025}}. In our case, at $T = 106\;$K, we observe a resistance drop from 186 to $96\;\Omega$ upon application of a $2.1\;$V, $200\;$µs voltage pulse (see~Fig.~\ref{fig:2}e). Although the switching to the hidden state is apparent from the transport measurements, the short lifetime of this metastable phase at \mbox{$T > 50\;$K} prevents direct detection by techniques with longer acquisition times, such as ARPES \cite{vaskivskyiControllingMetaltoinsulatorRelaxation2015}.

\paragraph*{Angle-resolved photoemission spectroscopy:}  
Figure~\ref{fig:3} summarizes spatially-resolved ARPES measurements performed on the device shown in Fig.~\ref{fig:2}, using a photon energy of $74\;$eV and a beam spot focused to \mbox{$1.0\times1.5\;$µm$^2$}. Figure~\ref{fig:3}b shows representative angle-integrated and momentum-resolved spectra of this flake. In particular, the dispersion of the main \mbox{valence} bands, formed by S~$3p$ orbitals~\cite{hughesChargeDensityWaves1976}, is well resolved (see also Fig.~S3$\dag$). This is further validated by a constant energy contour at \mbox{$E-E_{\mathrm{F}} = -0.3\;$eV} with details and resolution comparable to those in studies on bulk \mbox{crystals \cite{clercLatticedistortionenhancedElectronphononCoupling2006,wangBandInsulatorMott2020, yangVisualizationChiralElectronic2022, maklarCoherentLightControl2023, qiTemperatureInducedReversible2024}}.

A spatially resolved Ta~$4f$ core-level scan with a step size of $1.5\;$µm is used to group the data points based on spectral \mbox{similarity} by the $k$-means clustering algorithm~\cite{jainDataClusteringReview1999} (see Fig.~\ref{fig:3}d). This unsupervised learning algorithm partitions the dataset into $k$~clusters by minimizing the intra-cluster variance. We determine the optimal~$k$ using the ``elbow'' method, where diminishing returns in \mbox{reducing} the within-cluster sum of squares are observed. We obtain $k = 3$ (see Fig.~S4$\dag$), where the three clusters correspond to distinct regions of the scanned area: bulk ($>30$ layers) and monolayer \mbox{1\textit{T}--TaS\textsubscript{2}} flake, as well as Au surface, agreeing well with the optical microscope image (see Fig.~\ref{fig:3}a and \ref{fig:3}d, as well as the overlaid image in Fig.~S5$\dag$). All angle-integrated spectra within a cluster are averaged to obtain the mean and standard deviation shown in Fig.~\ref{fig:3}e. A Shirley-type background~\cite{shirleyHighResolutionXRayPhotoemission1972} is subtracted to account for inelastic scattering (see Fig.~S6$\dag$). 

Within the bulk-like cluster, the Ta~$4f$ core levels exhibit a splitting stemming from the different electronic environments of Ta~atoms within the star-of-David-shaped lattice distortion in the CDW~phase~\cite{wilsonChargedensityWavesSuperlattices1975}. As shown in the inset in Fig.~\ref{fig:3}e, there are three distinct Ta~sites, namely a central (A), as well as an \mbox{inner (B)} and outer (C) ring position. Due to their different electronic environment and the charge redistribution towards the central A~site, the Ta~$4f_{7/2}$ core levels appear at different binding energies, \mbox{\textit{i.e.} \mbox{$E-E_\mathrm{F}= -23.7\;$eV (C)}, \mbox{$-23.1\;$eV (B)}, and \mbox{$-23.0\;$eV} (A)} in accordance with measurements of bulk single-crystals \cite{wangDualisticInsulatorStates2024}. The ratio of the integrated intensities matches that of the Ta atoms of each species, \mbox{\textit{i.e.} A$\,:\,$B$\,:\,$C\;$ \widehat{=} $\;1$\,:\,$6$\,:\, $6}. We attribute deviations of the fit from the experimental data, most pronounced for the \mbox{C species}, to inequivalent \mbox{C sites} due to the complex out-of-plane stacking \cite{butlerMottnessUnitcellDoubling2020, huaEffectInterlayerStacking2025}, or puckering distortions \cite{bozinCrystallizationPolaronsCharge2023} present in the CDW states of \mbox{1\textit{T}--TaS\textsubscript{2}}. The discrepancy at the edge of the spectrum can be attributed to systematic errors from the background subtraction.

In contrast, the spectrum taken from the monolayer cluster shows a reduction of the splitting between the B and C species, alongside a reduced relative C intensity, which we attribute to the absence of out-of-plane stacking. A shift of \mbox{$\approx 0.1\;$eV} towards lower binding energies is observed, which we ascribe to charge transfer from Au to \mbox{1\textit{T}--TaS\textsubscript{2}} and is consistent with findings in monolayer MoS\textsubscript{2} on Au contacts \cite{velickyMechanismGoldAssistedExfoliation2018}. To study the influence of \mbox{1\textit{T}--TaS\textsubscript{2}}$\,$/$\,$Au interface in more detail, one could aim for a device with a \mbox{1\textit{T}--TaS\textsubscript{2}} monolayer that extends over the gapped region of the substrate in a future experiment, allowing for a direct comparison of the Au-bound and freestanding monolayer.


\section{Conclusions}

We have developed a resist-free device fabrication method tailored for exfoliation of vdW materials, combining contact patterning through a micro-stencil with gold-assisted exfoliation under UHV conditions. This approach enables well-defined device geometries with micron-sized features, including electrical connections for transport measurements and switching. The ability to fabricate devices under ambient conditions while performing exfoliation in UHV makes the approach particularly suitable for use with surface-sensitive techniques. As a proof of principle, we fabricate two-terminal devices from \mbox{1\textit{T}--TaS\textsubscript{2}} flakes, showcasing the feasibility of simultaneous electrical and ARPES measurements. We find that electrical contacts to the flakes are of high quality with contact resistances of only $\sim 10\; \Omega$. Spatially-resolved micro-beam ARPES demonstrates that the our device fabrication yields high-quality surfaces across \mbox{$\sim$100-µm} sized flakes with spectroscopic detail comparable to traditional measurements on bulk crystals. Our work presents a platform for studying \mbox{devices} made from vdW materials down to the single layer regime and enables combined transport and spectroscopic investigations that so far have not been permitted by other fabrication methods.\\


\section{Experimental methods}
\label{sec:experimental}

\paragraph*{Material synthesis:} 
Single crystals of \mbox{1\textit{T}--TaS\textsubscript{2}} are grown by chemical vapor transport using I$_2$ as a transport agent, following previously established procedures \cite{stojchevskaUltrafastSwitchingStable2014}.

\paragraph*{Shadow mask fabrication:} 
Stencils are fabricated from \mbox{250-µm} thick Si wafers utilizing laser lithography and DRIE. The electrode design is exposed on a \mbox{9-µm thick} film of positive tone photoresist (SPR~220~7.0, micro resist technology) using a direct-write laser lithography tool (DWL 66+, Heidelberg Instruments Mikrotechnik) at $405\;$nm. The Si~wafer is etched through in a DRIE tool (Omega Rapier \mbox{200-mm} process module, SPTS Technologies) using a variation of the highly anisotropic ``Bosch process'' optimized for low sidewall roughness and micron-scale features. It consists of alternating etching and passivation cycles. The combination of physical and chemical mechanisms enables highly directional etching through the entire wafer, while minimizing lateral etching and ensuring vertical sidewalls. 

\begin{figure*}
\centering
  \includegraphics[width=1\linewidth, trim=0cm 12.52cm 2.46cm 2.05cm, clip]{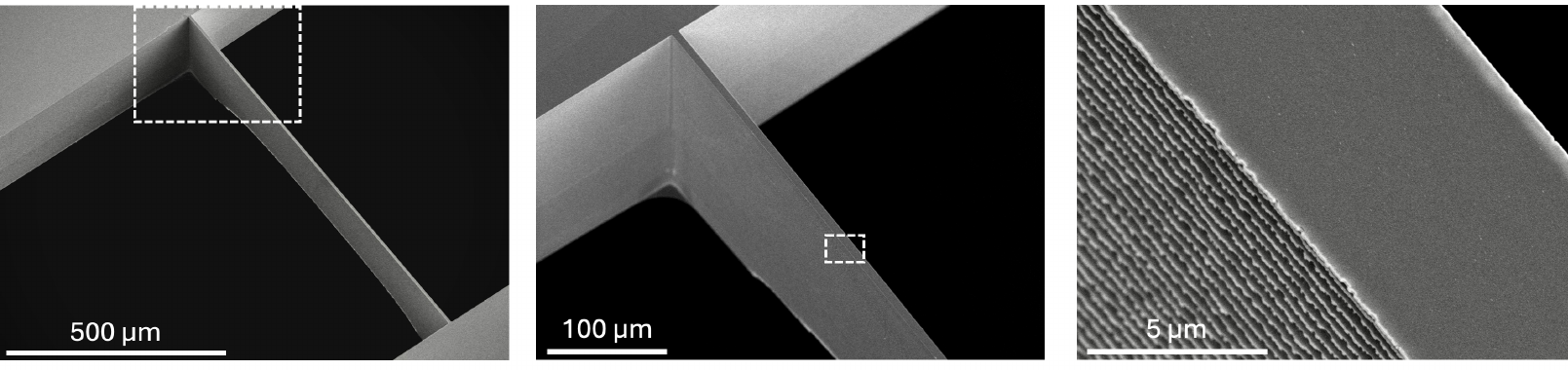}
  \caption{Scanning electron microscopy images of the shadow mask with a 7-µm wide, 1-mm long Si bridge at different magnifications (enlargements of the dashed boxes are shown on the right). The image on the right reveals the steep sidewalls with a characteristic scalloped profile, arising from the alternating passivation and etching cycles during the Bosch DRIE process.}
  \label{fig:4}
\end{figure*}

\noindent Our process is carried out at $0\,^\circ$C with the electrostatic chuck biased at~$6000\;$V and the coil current set to $10\;$A. The sequence of three steps (see the process parameters listed in Tab.~\ref{tab:drie_params}) starts with \textit{Si~etching} from the reaction between F radicals, from SF\textsubscript{6} plasma, and Si that results in volatile SiF\textsubscript{4} byproducts. Additional platen power is applied to accelerate plasma-generated ions toward the surface and remove material selectively. The chemical nature of this process leads to unwanted lateral etching. To counteract this effect, our process incorporates a \textit{passivation} step in which isotropic C\textsubscript{4}F\textsubscript{8} plasma deposits a fluorocarbon polymer layer on both the surface and the sidewalls of the etched trenches. This is followed by an intermediate \textit{base etching} step using a gaseous mixture of SF\textsubscript{6} and O$_2$. Platen bias power is applied to enhance anisotropy and remove most of the passivation layer at the base of the trenches, while preserving it on the sidewalls to protect them during the subsequent Si etch which continues until the passivation layer is completely consumed. These steps are repeated cyclically, where each cycle (see Fig.~\ref{fig:4}) removes a controlled amount of Si and prevents lateral etching, enabling the formation of high-aspect-ratio structures.

\begin{table}[t]
    \small
    \caption{Parameters of the Bosch DRIE process.}
    \centering
    \renewcommand{\arraystretch}{1.3}
    \setlength{\tabcolsep}{9.4pt}
    \begin{tabular}{lccc}
        \toprule
        \textbf{Process step} & \textbf{Passivation} & \textbf{Base etch} & \textbf{Si etch} \\ 
        \midrule
        Etch time [s] & 1.8 & 1.1 & 4.0 \\
        Pressure [mTorr] & 40 & 20 & 30 \\
        Primary source [W] & 2500 & 2500 & 2500 \\
        Secondary source [W] & 0 & 400 & 400 \\
        Platen power [W] & 0 & 50 & 50 \\
        \midrule
        \multicolumn{4}{l}{\textbf{Gas flows [sccm]}} \\
        \midrule
        SF$_6$ & 1 & 120 & 400 \\
        C$_4$F$_8$ & 300 & 1 & 1 \\
        O$_2$ & 1 & 120 & 1 \\
        \bottomrule
    \end{tabular}
    \label{tab:drie_params}
\end{table}

\paragraph*{Device design:} 
The geometry of the shadow mask used for this proof-of-principle study, featuring a $1\;$mm$~\times~7\;$µm bridge, is designed to form two-terminal devices where exfoliated flakes cover the predefined contact gaps. The contact dimensions are optimized for the typical size and density of exfoliated flakes while ensuring compatibility with the spatial resolution of the micro-beam ARPES setup. A gap size of $4-8\;$µm is found to provide a balance between the ability to resolve the area between contact pads and maintaining a high probability of flakes bridging. The vertical gap length~($1\;$mm) determines the average amount of flakes covering the gap---longer gaps increase the probability of multiple flakes, while shorter ones are preferred for single-flake transport measurements (at the risk of having no flake bridging the gap). The large lateral extent of the contact pads to the sides ($>1\;$mm) ensures sufficient exposed area next to the tape loop for the fixation of wires to establish electrical contact to the device.

\paragraph*{Contact deposition:} 
Ti/Au contacts are deposited on the substrate using the shadow mask in an electron beam evaporator (BAK Uni, Evatec) at pressures below $1 \times 10^{-6}\;$mbar. \mbox{A 2-nm Ti}~layer is deposited first to promote adhesion between SiO\textsubscript{2} and the \mbox{13-nm Au film~\cite{todeschiniInfluenceTiCr2017}}. The thickness of the Au layer is optimized to balance surface smoothness for exfoliation and mechanical durability during electrical pulsing.

\paragraph*{Substrates:} 
The devices are fabricated on 4-inch Si/SiO\textsubscript{2} wafers (MicroChemicals) with a thickness of $525 \pm 20\;$µm, p-type doping, a resistivity of $1-5$~m$\Omega$cm, bow/warp $<30\;$µm, and \mbox{$290\;$nm} of~SiO\textsubscript{2}.

\paragraph*{Material transfer:} 
After contact deposition, bulk \mbox{1\textit{T}--TaS\textsubscript{2}} crystals with typical dimensions of $1-3\;$mm in diameter and $\approx100\;$µm thickness are cleaved on tape loops, brought into contact with the substrate---aligning the bulk material by eye with the pre-patterned contacts---and gently pressed on to avoid trapping air pockets under the tape. Among other commonly used tapes, \textit{e.g.}~Nitto Blue Tape or Scotch Tape, we achieved the most reproducible results with silicone adhesive, UHV compatible Kapton tape. 
The exfoliation process strongly depends on the cleanliness and smoothness of the Au surface as contamination and roughness diminish the adhesion between the vdW material and Au, leading to a lower exfoliation yield~\cite{velickyMechanismGoldAssistedExfoliation2018}. This occurs because airborne organic contaminants accumulate on the Au~surface, turning it more hydrophobic and weakening the adhesion~\cite{smithHydrophilicNatureClean1980}. To minimize these effects, the transfer is performed rapidly after venting the evaporation chamber with N$_2$. Material transfer within $5\;$min consistently yields large-area flakes down to the monolayer limit, whereas longer time spans result in predominantly thicker ($\approx50-100\;$nm) flakes with lower yield.

\paragraph*{Device assembly:}
The Si wafer containing 24~devices is cleaved manually into dies and glued on standard flag-style copper plates (see Fig.~S1$\dag$) using Ag epoxy (EPO-TEK H20E, Epoxy \mbox{Technology}). Electrical connections to the device are also made using Ag epoxy by attaching insulated copper wires to the Au~pads. The curing of the epoxy is performed under conditions that minimize thermal stress: We heat the fully assembled sample holder to the lowest temperature specified 
that ensures full curing of the epoxy, \textit{i.e.} $80\,^\circ$C, thereby avoiding thermal expansion that could strain the vdW flake or introduce residues from melting the adhesive material of the tape.

\paragraph*{Electrical measurements:}
Measurements are carried out using a pulsed current source (\mbox{Keithley 6221}, Tektronix) and \mbox{nanovoltmeter} (Keithley 2182). Resistance values are obtained from a linear fit to the slope of \textit{IV} curves on a $\pm 10\;$µA current interval. Pulsed \textit{IV}~curves to measure the switching to the hidden state are recorded using a pulsed sweep mode, applying current pulses from 0 to 16$\;$mA in $100\;$µA steps. Each pulse has a width of $200\;$µs, but the voltage is measured only during the second half of the pulse, \textit{i.e.} after a $100\;$µs source delay. Between pulses, a pause of $100\;$ms minimizes heating.

\paragraph*{Angle-resolved photoemission spectroscopy:}
Measurements are conducted at the Spectromicroscopy endstation at Elettra Sincrotrone Trieste with a photon energy of $74\;$eV and an energy resolution of~$\approx45\;$meV. A beam footprint of \mbox{$1.0\times1.5\;$µm$^2$} is achieved through focusing with a zone plate and a Schwarzschild objective. Spectra are acquired using a hemispherical analyzer (MB Scientific AB A-1) equipped with a 2D~deflector system, \mbox{accepting} a $35^\circ$ cone. Devices are cleaved at room \mbox{temperature} in a preparation chamber~($<1\times 10^{-9}\;$mbar pressures) and transferred to the manipulator head in the main chamber~($<7\times10^{-10}\;$mbar). The cooling rate is controlled at~$\approx(1$–$2)\;$K/min to cross the \mbox{NC-CCDW} transition of 1\textit{T}--TaS$_2$ in a controlled manner. The binding energy axis is aligned to $E_{\mathrm{F}}$ by fitting the angle-integrated spectrum with a Fermi-Dirac distribution. 

\paragraph*{Core-level analysis:}
To analyze the Ta 4$f$$_{7/2}$ core-level spectra, we employ a multi-step fitting procedure. First, we perform spatial clustering using a $k$-means algorithm~\cite{jainDataClusteringReview1999} to group different regions of interest based on spectral similarity. Within each cluster, all angle-integrated spectra are averaged to obtain the cluster mean and standard deviation. A Shirley-type background~\cite{shirleyHighResolutionXRayPhotoemission1972} is subtracted to account for inelastic scattering contributions. The core-level spectra are fitted using three Voigt profiles, corresponding to the three in-equivalent Ta sites of the star-of-David pattern forming the CDW state. The Voigt function, defined as the convolution of a Lorentzian and a Gaussian profile, accounts for both the intrinsic lifetime broadening of the core-hole state (Lorentzian) and the instrumental resolution of the measurement setup (Gaussian). The initial peak positions were set to $-23.64\;$eV, $-23.10\;$eV, and $-22.97\;$eV, based on prior studies on bulk crystals~\cite{wangDualisticInsulatorStates2024}, and allowed to vary within $\pm 0.2\;$eV. The initial amplitude ratios were set to 1:6:6 (A:B:C), and the Gaussian and Lorentzian widths constrained to $\sigma \leq 0.2$ and $\gamma \leq 0.2$, respectively. The best fit parameters were determined using least-squares optimization. Intensity ratios were established by integrating the peak areas.

\paragraph*{Atomic force microscopy:}
Measurements are conducted after the ARPES experiment using a Dimension 3100 system (Bruker) to establish the thickness and homogeneity of the exfoliated \mbox{1\textit{T}--TaS\textsubscript{2}} flakes. Line profiles were extracted across the flake edges from full-area scans. The resulting profiles (see~Fig.~S2$\dag$), were leveled by subtracting the local slope and fitted to determine the step height, which corresponds to the flake thickness.

\paragraph*{Scanning electron microscopy:}
High-resolution images of the shadow mask were obtained using a scanning electron microscope (Regulus 8230, Hitachi).

\section*{Author contributions}

N.T., C.B., R.V. and I.G. conceived the project with input from D.K., Y.E., D.M. and S.G. 
N.T., C.B. and R.V. fabricated the devices. 
I.G. fabricated the shadow mask with input from D.K. 
N.T., C.B., R.V., E.D.V, A.M., Y.V., J.L. and A.B. carried out the ARPES experiments. 
N.T. analyzed the ARPES data with input from S.E. 
N.T., C.B., R.V., I.G., D.M. and S.G. wrote the manuscript with input from all co-authors.

\section*{Data availability} 

Data of this article, including ARPES and AFM scans, transport data and photographs are available at the Zenodo repository under \url{https://doi.org/10.5281/zenodo.15482370}.

\section*{Conflict of Interest}

The authors declare no competing interest.


\section*{Acknowledgments} 

We thank the PSI~PICO operations team for technical support. The authors acknowledge Elettra Sincrotrone Trieste for providing access to its synchrotron radiation facilities at the Spectromicroscopy beamline. We also thank M. D. Watson for technical support during preliminary measurements at the I05~beamline at Diamond Light Source. This research was funded by the Swiss National Science Foundation~(SNSF) and the Slovenian Research And Innovation Agency~(ARIS) as a part of the WEAVE framework Grant Number~213148 (ARIS Project N1-0290). C.B. and S.A.E. acknowledge funding from the European Research Council under the European Union’s Horizon 2020 Research and Innovation
Programme, within Grant Agreement~810451 (HERO). R.V., A.M., Y.V and D.M. thank ARIS for funding the research program P1-0040, and D.M. thanks ARIS for funding the 
research program~\mbox{J7-3146}. \\

\balance

\bibliography{bibliography}
\bibliographystyle{rsc} 

\includepdf[pages=-]{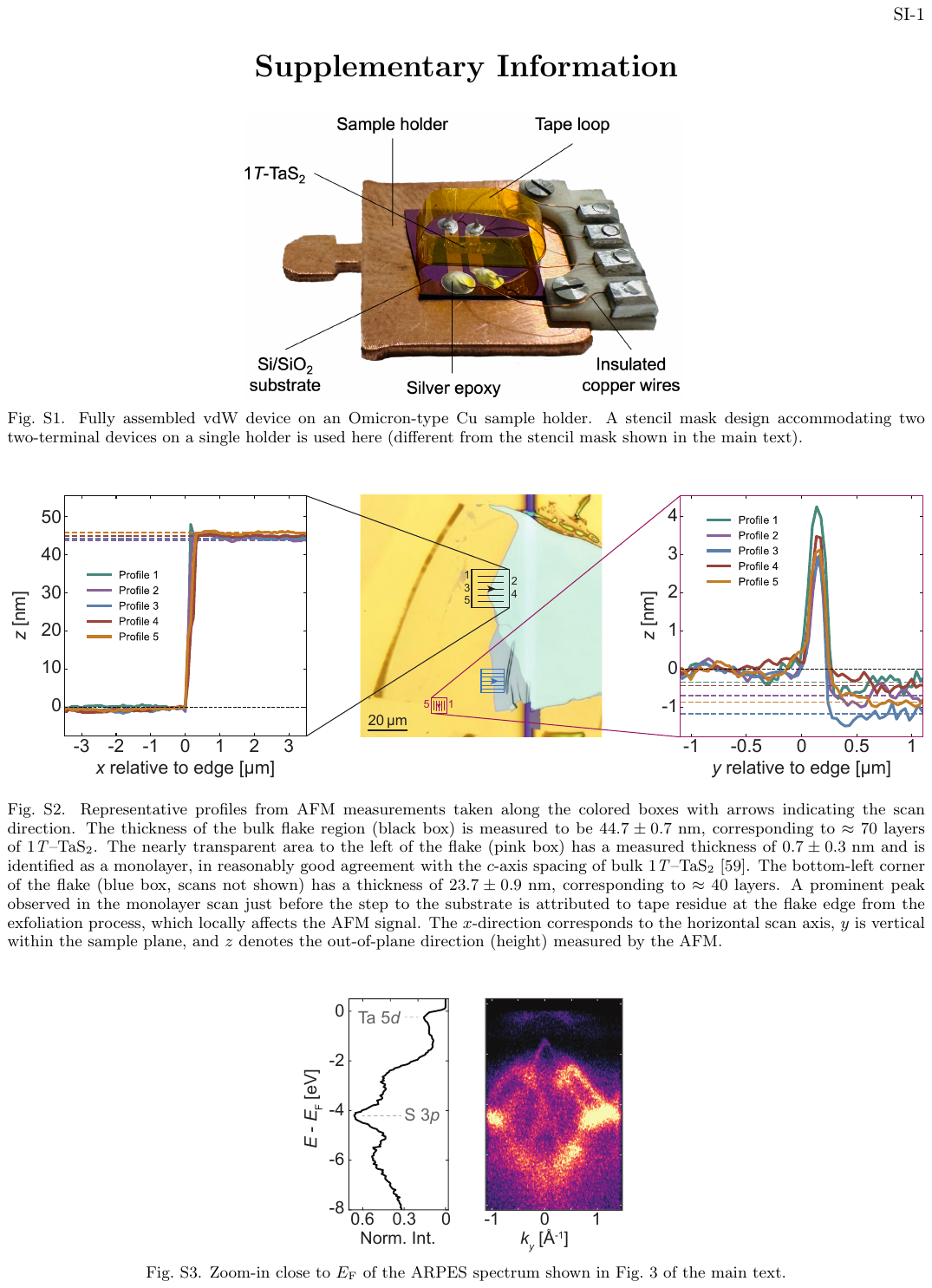}

\end{document}